\documentclass{ptephy_v1}

\preprintnumber{XXXX-XXXX} 

\usepackage{braket}
\usepackage{graphicx}

\begin{document}

\title{Alignment of wave functions for angular momentum projection}


\author{Yasutaka Taniguchi}
\affil{Faculty of Health Sciences, Nihon Institute of Medical Science, 1276 Shimogawara, Moroyama, Saitama 350-0435, Japan\email{yasutaka@nims.ac.jp}}

%
%


\begin{abstract}%
Angular momentum projection is used to obtain eigen states of angular momentum from general wave functions.
Multi-configuration mixing calculation with angular momentum projection is an important microscopic method in nuclear physics.
For accurate multi-configuration mixing calculation with angular momentum projection, concentrated distribution of $z$ components $K$ of angular momentum in the body-fixed frame  ($K$-distribution) is favored.
Orientation of wave functions strongly affects $K$-distribution.
Minimization of variance of $\hat{J}_z$ is proposed as an alignment method to obtain wave functions that have concentrated $K$-distribution.
Benchmark calculations are performed for $\alpha$-$^{24}$Mg cluster structure, triaxially superdeformed states in $^{40}$Ar, and Hartree-Fock states of some nuclei.
The proposed alignment method is useful and works well for various wave functions to obtain concentrated $K$-distribution.
\end{abstract}

\subjectindex{xxxx, xxx}

\maketitle

\section{Introduction}

Symmetry is important to investigate natures of systems.
Hamiltonian has rotational symmetry, which is critical for isolated systems such as nuclei.
Static states of nuclei are eigen states of angular momentum.
In order to study nuclear properties theoretically in details, multi-configuration mixing calculation of eigen states of angular momentum are performed such as few-body calculations\cite{PhysRevC.64.044001},  Monte Carlo shell model (MCSM)\cite{OTSUKA2001319}, Hartree-Fock + generator coordinate method (HF-GCM)\cite{RevModPhys.75.121}, fermionic molecular dynamics (FMD)\cite{ROTH20043}, and antisymmetrized molecular dynamics (AMD)\cite{Kanada-En'yo01012012}.
The first-principle calculations in which all symmetries are taken into account are performed in only few-body systems in nuclear physics.
Simple wave functions such as Slater determinants are often used as basis wave functions for many-body systems.

In MCSM, HF-GCM, FMD, and AMD, Slater determinants with angular momentum projection are used as basis wave functions of multi-configuration mixing calculation.
A Slater determinant is a simple and useful wave function to describe various deformation but it can violate rotational and other symmetries.
In order to restore rotational symmetry, angular momentum projection is performed.
By multi-configuration mixing calculation with angular momentum projection, detailed nuclear structure can be clarified with taking into account many-body correlations.
In the calculation, basis wave functions that contain fewer number of $K$-components are favored for accurate calculations, where $K$ is a $z$ component of total angular momentum in the body-fixed frame.
It is because different $K$-components are superposed in the calculation, which is called $K$-mixing, and the different $K$-components are nonorthogonal in general.
Mixing of nonorthogonal components can occur cancellation of digits, which lose accuracy.

Distribution of $K$-components ($K$-distribution) strongly depends on orientation of wave functions.
Suppose an axially symmetric wave function.
When the symmetric axis is set to $z$-axis, the wave function is invariant for rotation around $z$-axis, which shows the wave function contains only $K = 0$ components. 
On the other hand, when the symmetric axis is set to other direction, the wave function is variant for rotation around $z$-axis and contains many kinds of $K$-components although the wave function is symmetric.
Wave functions with different orientations have different $K$-distribution, but they give same energy after angular momentum projection with $K$-mixing\cite{bender:024309,PhysRevC.79.044312,PhysRevC.81.064323} in principle, but concentrated $K$-distribution is better for accuracy of numerical calculation.

One choice to align wave functions is diagonalization of moment-of-inertia matrix.
If a wave functions is axially symmetric, the symmetric axis should be the long or short principal axis of inertia for prolate or oblate shapes, respectively.
This choice seems to work well for deformed nuclei, but it is unclear whether it works well for exotic structures such as cluster structures and triaxially deformed structures, and nuclei that are spherical but have finite spin such as odd and odd-odd nuclei.
A general method of alignment that can apply to various nuclei is required.

In this paper, a method to align wave functions to concentrate $K$-distribution to fewer components is proposed.
By minimization of variance of $\hat{J}_z$, which is $z$ component of angular momentum, wave functions are aligned and $K$-distribution is concentrated to fewer components.
It is shown that this method works well for spherical and deformed nuclei including exotic deformation such as cluster structures and triaxially deformed structures.

This paper is organized as follows.
In Sec.~2, formalism of the method and model wave functions are explained.
Numerical results are shown in Sec.~3.
In Sec.~4, $K$-distribution of wave functions aligned by the proposed method and a referenced method is compared.
Finally, conclusions are given in Sec.~5.

\section{Formalism}

\subsection{Angular momentum projection}

The angular momentum projection operator $\hat{P}_{MK}^J$ onto $\ket{J M}$ state from $\ket{J K}$ state\cite{PhysRev.95.122} is given as
\begin{equation}
 \hat{P}_{MK}^J = \frac{2J + 1}{8\pi^2}\int D_{MK}^{J\ast} (\alpha, \beta, \gamma) \hat{R}(\alpha, \beta, \gamma) \mathrm{d}\alpha \mathrm{d}\beta \mathrm{d}\gamma,
\end{equation}
where $J$ is total angular momentum, $M$ and $K$ are $z$-component of angular momentum in laboratory and body-fixed frames, respectively, $\alpha$, $\beta$, and $\gamma$ are Euler angles, and $\hat{R}$ is rotation operator defined as
\begin{equation}
 \hat{R}(\alpha, \beta, \gamma) = e^{-i\alpha \hat{J}_z} e^{-i\beta \hat{J}_y} e^{-i\gamma \hat{J}_z},
\end{equation}
where $\hat{\mathbf{J}}$ is angular momentum operator.
$D^{J}_{MK}(\alpha, \beta, \gamma)$ is the Wigner function given as
\begin{equation}
 D^J_{MK}(\alpha, \beta, \gamma) = e^{-i M \alpha} d^J_{MK} (\beta) e^{-i K \gamma},
\end{equation}
here $ d^J_{MK} (\beta)$ is Wigner's formula.
The angular momentum projection operator is divided into three parts as,
\begin{equation}
 \hat{P}_{MK}^J = (2J + 1) \hat{P}_M^{(z)} \hat{P}_{MK}^{(y)J} \hat{P}_K^{(z)},\label{PPzPyPz}
\end{equation}
where
\begin{eqnarray}
 \hat{P}_K^{(z)} &=& \frac{1}{2\pi} \int_0^{2\pi} \mathrm{d}\phi e^{i(K - \hat{J}_z) \phi},\\
 \hat{P}_{MK}^{(y)J}&= & \frac{1}{2} \int_{-1}^1 \mathrm{d}(\cos\theta) d_{MK}^{J} (\theta) e^{-i \hat{J}_y \theta},
\end{eqnarray}
and $\hat{P}_K^{(z)}$ is called $K$-projection operator.
Using the angular momentum projection operator, a resultant wave function $\ket{\Psi_{J_nM}}$ of multi-configuration mixing calculation for the $n$th $J$ state is given as
\begin{equation}
 \ket{\Psi_{J_n M}} = \sum_{ij} \hat{P}_{MK_i} \ket{\Phi_j} f_{ij}^{J_n} = (2J + 1) \hat{P}_M^{(z)} \sum_{ij} \hat{P}_{M K_i}^{(y)} \ket{\hat{P}_{K_i}^{(z)} \Phi_j} f_{ij}^{J_n},\label{MCM}
\end{equation}
where $\ket{\Phi_j}$ is a Slater determinant for a basis wave function of the multi-configuration mixing calculation, and $f_{ij}^J$ is a weight of the each basis wave function, which is determined by diagonalization of Hamiltonian and norm matrices as
\begin{equation}
 \braket{\Psi_{J_m M} | 
  \left(
   \begin{array}{c}
    \hat{H}\\
    1\\
   \end{array}
   \right)
  | \Psi_{J_n M} }
  =
  \left(
   \begin{array}{c}
    E_{J_n}\\
    1 \\
   \end{array}
   \right)
  \delta_{mn},
\end{equation}
here $E_{J_n}$ is an expectation value of energy of the $\ket{\Psi_{J_n M}}$ state.
As seen in right hand side of Eq.~(\ref{MCM}), a $K$-projected wave function $\ket{\hat{P}_{K_i}^{(z)} \Phi_j}$ is treated as a basis wave function.
If $K_i$-component of $\ket{\Phi_j}$ is enough small as,
\begin{equation}
 \hat{P}_{K_i}^{(z)} \ket{\Phi_j} \sim 0,
\end{equation}
the wave function $\ket{\hat{P}_{K_i}^{(z)} \Phi_j}$ can be eliminated from the basis set.

Different $K$-components of a wave function are nonorthogonal, which suppress accuracy of diagonalization of Hamiltonian and norm matrices because of cancellation.
More concentrated $K$-distribution is better for multi-configuration mixing calculation.
Orientation of wave functions strongly depends on $K$-distribution.
Effective alignment of the original wave function,
\begin{equation}
 \ket{\Phi_i} \rightarrow \ket{\Phi_i^\prime} = \hat{R} \ket{\Phi_i},\label{align}
\end{equation}
is required for accurate multi-configuration mixing calculation after angular momentum projection.

\subsection{Alignment of wave functions}

In this paper, a method of alignment by minimization of variance $V_{zz}^{J}$ of $\hat{J}_z$ is proposed.
$V_{zz}^{J}$ is a $(z, z)$ component of variance matrix $\mathsf{V}^J$ of angular momentum, and a $(\sigma, \rho)$ component of the matrix $\mathsf{V}^J$ is defined as
\begin{equation}
 V_{\sigma\rho}^J = \Braket{\frac{\hat{J}_\sigma \hat{J}_\rho + \hat{J}_\rho \hat{J}_\sigma}{2}} - \braket{\hat{J}_\sigma} \braket{\hat{J}_\rho},
\end{equation}
here $\sigma$ and $\rho$ are $x$, $y$, or $z$.

By alignment of the original wave function, Eq.~(\ref{align}), the variance matrix $\mathsf{V}^J$ is transformed as
\begin{equation}
 \mathsf{V}^J \rightarrow \mathsf{V}^{J\prime} = \mathsf{R}(\alpha, \beta, \gamma) \mathsf{V}^J {}^\mathrm{t}\mathsf{R} (\alpha, \beta, \gamma),
\end{equation}
where $\mathsf{R} (\alpha, \beta, \gamma)$ is a rotation matrix for Euler angles, $\alpha, \beta, \gamma$ as
\begin{eqnarray}
 &&\mathsf{R}(\alpha, \beta, \gamma) \nonumber\\
 &=& 
  \left(
   \begin{array}{rrr}
    \cos\alpha \cos\beta \cos\gamma - \sin\alpha \sin\gamma & -\cos\alpha \cos\beta \sin\gamma - \sin\alpha \cos\gamma & \cos\alpha \sin\beta \\
    \sin\alpha \cos\beta \cos\gamma + \cos\alpha \sin\gamma & -\sin\alpha \cos\beta \sin\gamma + \cos\alpha \cos\gamma & \sin\alpha \sin\beta\\
    -\sin\beta \cos\gamma & \sin\beta \sin\gamma & \cos\beta\\
   \end{array}
  \right).\nonumber\\
\end{eqnarray}
Optimized Euler angles are obtained by diagonalization of the variance matrix $\mathsf{V}^J$ because it is a real symmetric matrix.

For comparison, $K$-distribution of wave functions aligned by diagonalization of moment-of-inertia matrix $\mathsf{I}$ is also calculated.
The $(\sigma, \rho)$ element $I_{\sigma \rho}$ of the moment-of-inertia matrix $\mathsf{I}$ is
\begin{equation}
 I_{\sigma \rho} = \sum_i \braket{\hat{r}_{i\sigma} \hat{r}_{i\rho}},
\end{equation}
where $\hat{r}_{i\sigma}$ is a $\sigma$ component of the coordinate operator for the $i$th particle.
$z$-axis is set to long (short) principle axis of inertia for the case of quadrupole deformation parameter $\gamma_2 < 30^\circ$ ($\gamma_2 > 30^\circ$).

\subsection{Model wave functions}

Deformed-basis AMD wave functions\cite{PhysRevC.69.044319} are used as model wave functions of benchmark calculations.
A deformed-basis AMD wave function $\ket{\Psi}$ is defined as
\begin{eqnarray}
 \ket{\Psi} &=& \hat{\mathcal{A}} \ket{\psi_1, \psi_2, \cdots, \psi_A},
\end{eqnarray}
where $\hat{\mathcal{A}}$ is antisymmetrization operator, $A$ is the mass number, and $\ket{\psi_i}~(i = 1, ..., A)$ is a single-particle wave function.
The single-particle wave function $\ket{\psi_i}$ is defined as
\begin{eqnarray}
 \ket{\psi_i} &=& \ket{\phi_i \otimes \sigma_i \otimes \tau_i},\\
 \braket{\mathbf{r} | \phi_i} &= & \left( \frac{\det{\mathsf{M}}}{\pi^3} \right)^{\frac{1}{4}} \exp \left[ -\frac{1}{2} (\mathbf{r} - \mathbf{Z}_i)\cdot \mathsf{M}(\mathbf{r} - \mathbf{Z}_i) \right],\\
 \ket{\sigma_i}&= & \alpha_i \ket{\uparrow} + \beta_i \ket{\downarrow},\\
 \ket{\tau_i}&= & \ket{p}\ \mbox{or}\ \ket{n},
\end{eqnarray}
where $\ket{\phi_i}$, $\ket{\sigma_i}$, and $\ket{\tau_i}$ are spatial, spin, and isospin parts, respectively.
The spatial part $\ket{\phi_i}$ form a Gaussian wave packet, and $\mathsf{M}$ and $\mathbf{Z}_i$ denote width and center of the wave packet, respectively.
The width matrix $\mathsf{M}$ is a real $3\times 3$ symmetric matrix, which is common to all nucleons, and $\mathbf{Z}_i$ is a complex vector.
$\ket{\uparrow}$ and $\ket{\downarrow}$ are up and down spin, respectively, and $\alpha_i$ and $\beta_i$ are complex parameters to denote spin direction.
$\ket{p}$ and $\ket{n}$ are proton and neutron, respectively.
All parameters are determined by energy variational calculations.

\section{Results}

\subsection{Deformation of wave functions for benchmark calculations}

\begin{table}[tbp]
 \begin{center}
  \caption{
  Quadrupole deformation parameters $(\beta_2, \gamma_2)$ of wave functions used in benchmark calculations. 
  The first, second, and third columns show kinds of nuclides, configurations, and quadrupole deformation parameters of the wave functions, respectively.
  }
  \label{deformation}
  \begin{tabular}{ccc}
   \hline
   nuclide  & configuration      & $(\beta_2, \gamma_2)$\\
   \hline
   $^{12}$C & HF                 & $(0.25, 59.8^\circ)$\\
   $^{13}$C & HF                 & $(0.00, 19.0^\circ)$\\
   $^{14}$N & HF                 & $(0.02, ~1.0^\circ)$\\
   $^{15}$N & HF                 & $(0.00, ~9.0^\circ)$\\
   $^{16}$N & HF                 & $(0.09, 59.1^\circ)$\\
   $^{16}$O & HF                 & $(0.00, ~1.5^\circ)$\\
   $^{17}$O & HF                 & $(0.04, 57.0^\circ)$\\
   $^{18}$F & HF                 & $(0.12, 58.8^\circ)$\\
   $^{19}$F & HF                 & $(0.27, ~0.1^\circ)$\\
   $^{20}$F & HF                 & $(0.29, ~0.1^\circ)$\\
   $^{20}$Ne& HF                 & $(0.39, ~0.0^\circ)$\\
   $^{21}$Ne& HF                 & $(0.38, ~0.1^\circ)$\\
   $^{40}$Ar& triaxial SD        & $(0.48, 11.2^\circ)$\\
   $^{28}$Si& $\alpha$-$^{24}$Mg & $(0.49, 42.8^\circ)$\\
   \hline
  \end{tabular}
 \end{center}
\end{table}

As benchmark calculations, $K$-distribution of $\alpha$-$^{24}$Mg cluster structure of $^{28}$Si, Hartree-Fock (HF) states in $^{12,13}$C, $^{14,15,16}$N, $^{16,17}$O, $^{18, 19, 20}$F, and $^{20, 21}$Ne, and triaxially superdeformed (SD) state in $^{40}$Ar are calculated.
Those wave functions are obtained by energy variational calculations.
For $\alpha$-$^{24}$Mg cluster structure of $^{28}$Si and triaxially SD state of $^{40}$Ar, constraint potentials for intercluster distance\cite{PTP.112.475} between $\alpha$ and $^{24}$Mg and quadrupole deformation parameter $\beta_2$, respectively, are added to energy.
The Gogny D1S force is used as an effective interaction. 

Table~\ref{deformation} shows quadrupole deformation parameter $(\beta_2, \gamma_2)$ of wave functions used in benchmark calculations.
HF states of $^{13}$C, $^{14,15,16}$N,$^{16.17}$O have spherical shapes ($\beta_2 < 0.1$).
Other HF states are axially symmetric deformed shapes ($\gamma_2 \simeq 0^\circ$ or $60^\circ$).
The SD state in $^{40}$Ar and $\alpha$-$^{24}$Mg cluster structures in $^{28}$Si form triaxially deformed shapes ($\gamma_2 = 11.2^\circ$ and $42.8^\circ$, respectively).

\subsection{$K$-distribution}

\begin{figure}[tbp]
 \begin{center}
  \begin{tabular}{cc}
   \includegraphics[width=0.45\textwidth]{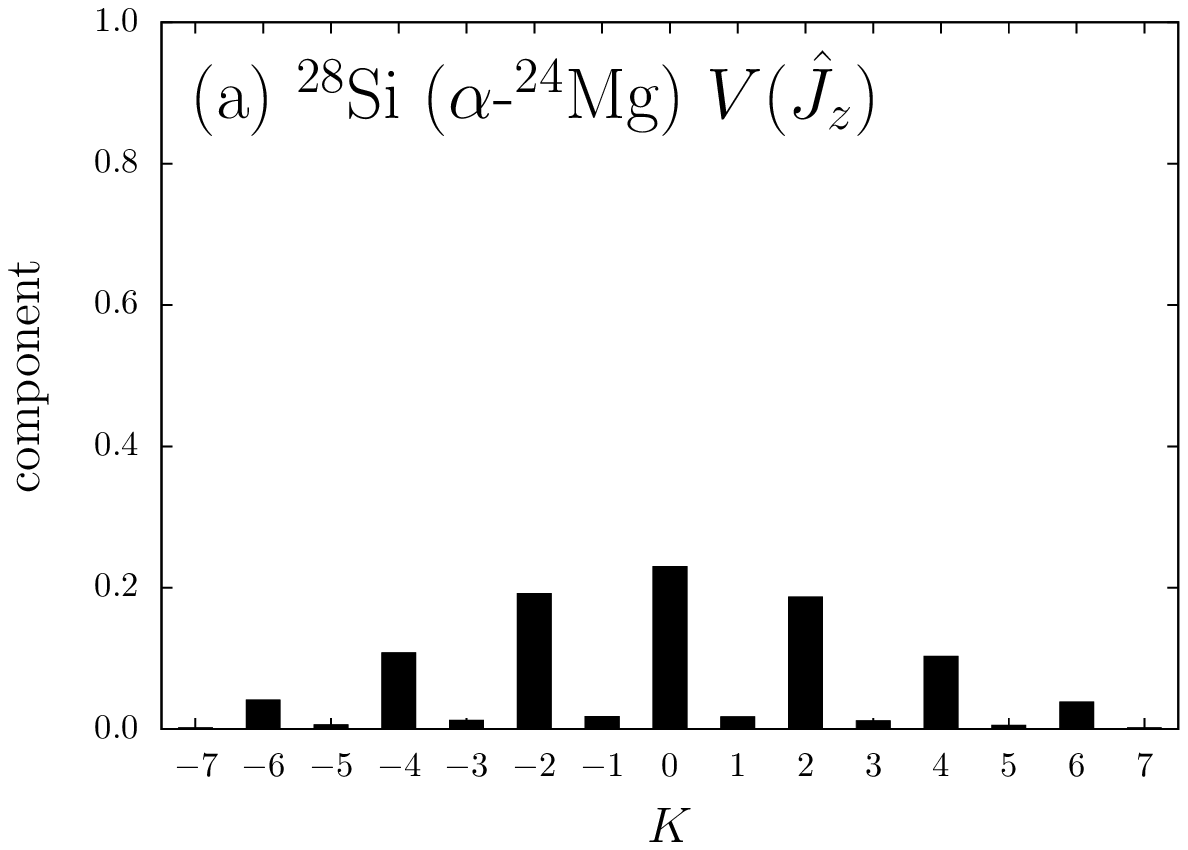}&
   \includegraphics[width=0.45\textwidth]{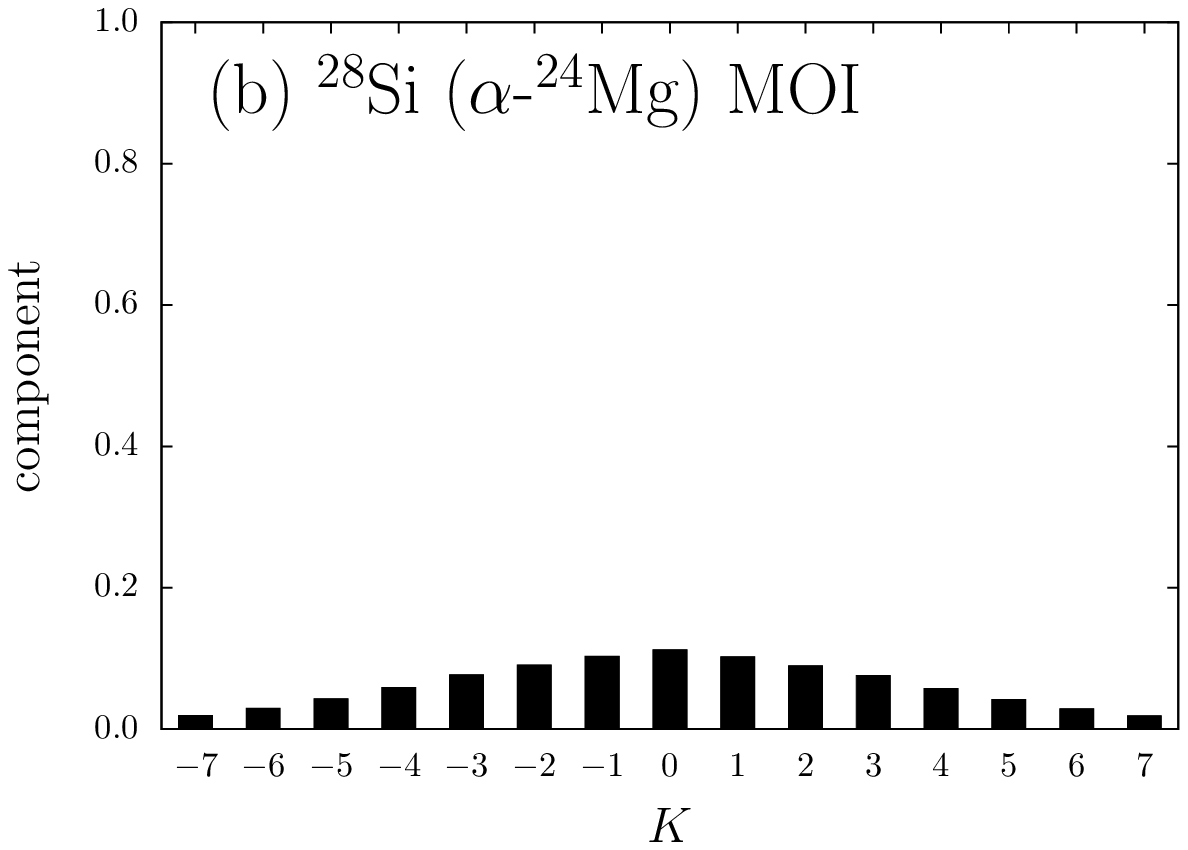}\\
  \end{tabular}
  \caption{
  $K$-distribution for $\alpha$-$^{24}$Mg cluster structure of $^{28}$Si.
  Wave functions are aligned by (a) minimization of variance of $\hat{J}_z$ and (b) diagonalization of the moment-of-inertia matrix.
  }
  \label{Si_K}
 \end{center}
\end{figure}


Figure~\ref{Si_K} shows $K$-distribution for $\alpha$-$^{24}$Mg cluster structure of $^{28}$Si.
Wave functions are aligned by minimization of variance of $\hat{J}_z$ and diagonalization of moment-of-inertia matrix.
The wave function aligned by minimizing variance of $\hat{J}_z$ contains even $K$-components dominantly [Fig.~\ref{Si_K}(a)].
On the other hand, in the wave function aligned by diagonalization of moment-of-inertia matrix, $K$-components are widely distributed over even and odd numbers [Fig.~\ref{Si_K}(b)].

\begin{figure}[tbp]
 \begin{center}
  \begin{tabular}{cc}
   \includegraphics[width=0.45\textwidth]{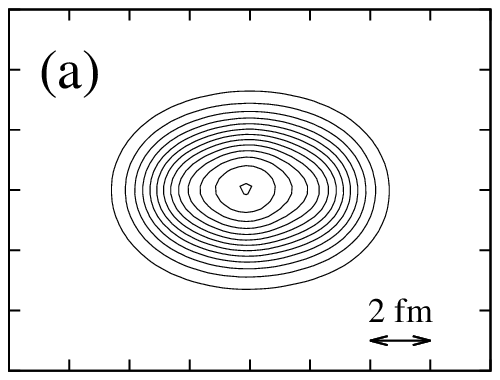}&
   \includegraphics[width=0.45\textwidth]{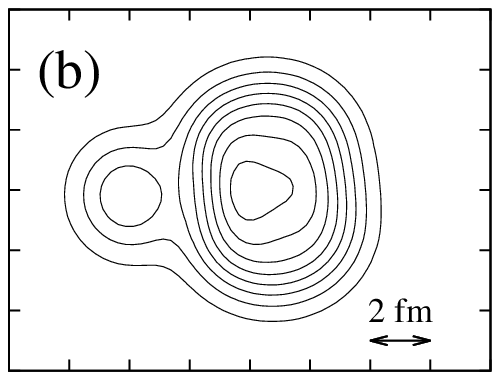}\\
  \end{tabular}
  \caption{
  Density distribution of $\alpha$-$^{24}$Mg cluster structure of $^{28}$Si integrated over $z$-axis on the $xy$-plane.
  Wave functions are aligned by (a) minimization of variance of $\hat{J}_z$ and (b) diagonalization of moment-of-inertia matrix.
  Contour lines are drawn every $0.1~\mathrm{fm^{-2}}$.
  }
  \label{28Si_density}
 \end{center}
\end{figure}

Figure~\ref{28Si_density} shows density distribution of $\alpha$-$^{24}$Mg cluster structure of $^{28}$Si.
Density is integrated over $z$-axis.
Wave functions are aligned by minimization of variance of $\hat{J}_z$ [Fig.~\ref{28Si_density}(a)] and diagonalization of moment-of-inertia matrix [Fig.~\ref{28Si_density}(b)].
The density distribution of Figs.~\ref{28Si_density}(a) and (b) are much different.
The density distribution in Fig.~\ref{28Si_density}(a) is symmetric for $\pi$ rotation around $z$-axis.
On the other hand, nucleons are separated into $\alpha$ (left) and $^{24}$Mg (right) clusters in Fig.~\ref{28Si_density}(b), and there is no symmetry for rotation around $z$-axis.
The differences are caused by choices of $z$-axis.
$^{24}$Mg cluster is deformed, and its long axis is orthogonal to the line that connects centers of mass of $\alpha$ and $^{24}$Mg clusters in the wave function,
and the lines that connect centers of mass of $\alpha$ and $^{24}$Mg clusters are parallel and orthogonal to $z$-axes for Figs.~\ref{28Si_density}(a) and (b), respectively.
The symmetric density distribution for $\pi$ rotation around $z$-axis in Fig.~\ref{28Si_density}(a) reflects even-$K$ dominance in Fig.~\ref{Si_K}(a).


\begin{figure}[tbp]
 \begin{center}
  \begin{tabular}{cc}
   \includegraphics[width=0.45\textwidth]{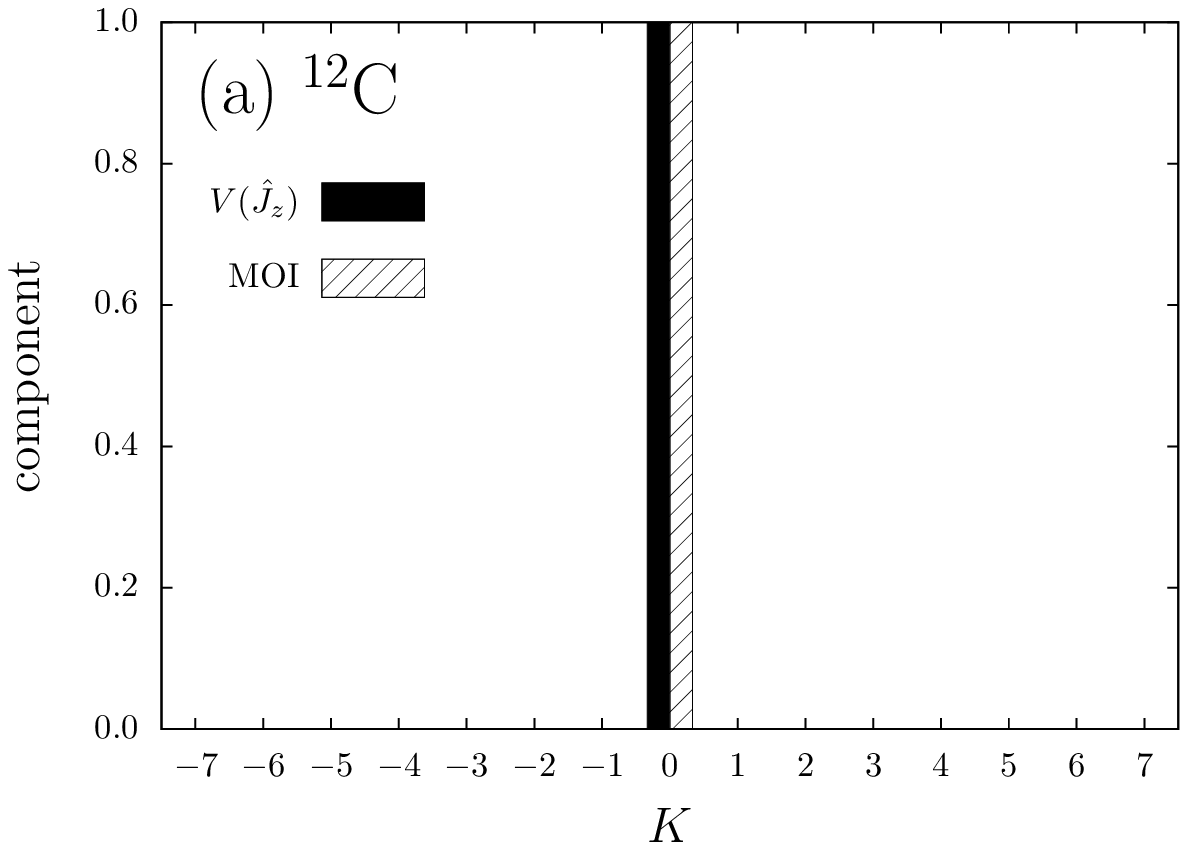}&
   \includegraphics[width=0.45\textwidth]{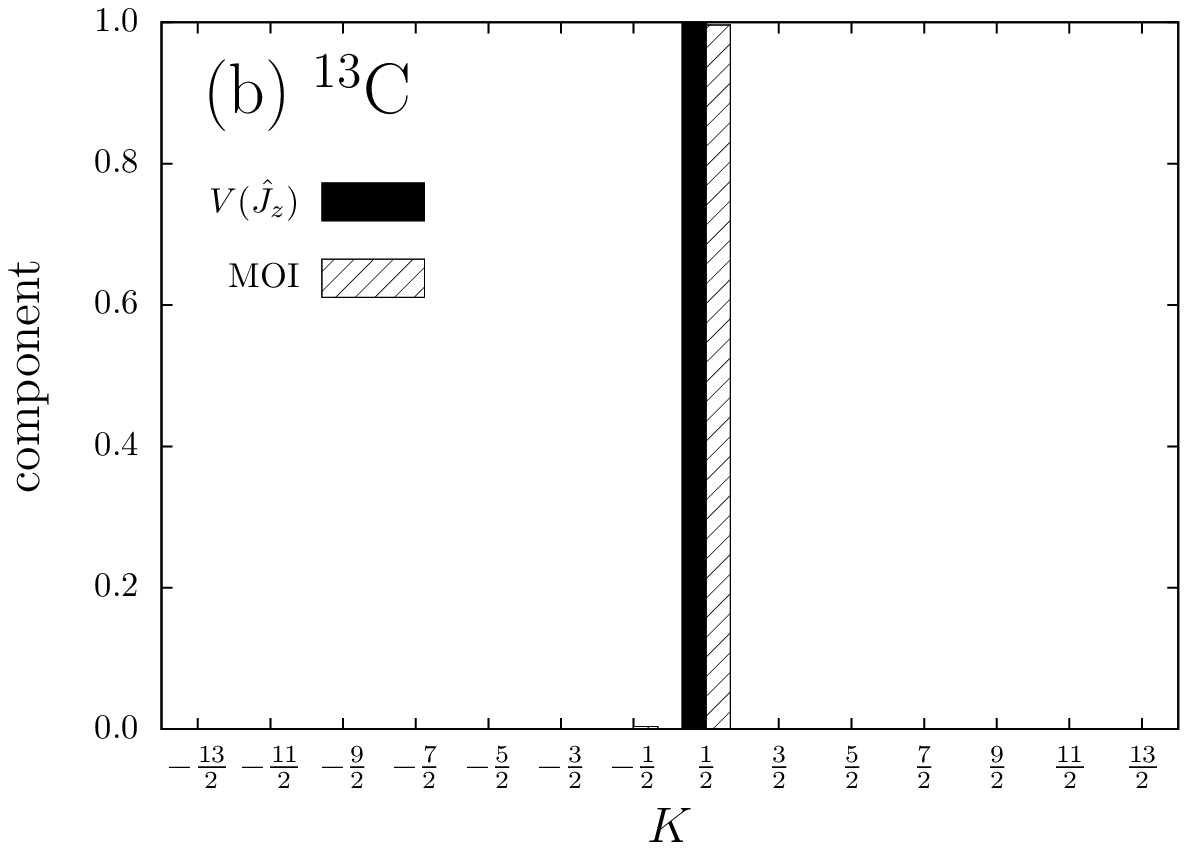}\\
  \end{tabular}
  \caption{
  $K$-distribution for HF states of (a)~$^{12}$C and (b)~$^{13}$C.
  Filled and slash-pattern bars show expectation values of $K$-projection operator for each $K$ number of wave functions aligned by minimization of variance of $\hat{J}_z$ and diagonalization of moment-of-inertia matrix, respectively.
  }
  \label{C_K}
 \end{center}
\end{figure}

Figure~\ref{C_K} shows $K$-distribution for HF states of $^{12,13}$C aligned by minimization of variance of $\hat{J}_z$ and diagonalization of moment-of-inertia matrix.
Both methods for alignment give similar results.
$^{12}$C and $^{13}$C contain only $K = 0$ and $\frac{1}{2}$ components, respectively, which shows axially symmetric form of the wave functions around $z$-axis.
The $K$-components of $^{12}$C and $^{13}$C reflect spin of their ground states that are $J^\pi = 0^+$ and $\frac{1}{2}^-$, respectively.

\begin{figure}[tbp]
 \begin{center}
  \begin{tabular}{cc}
   \includegraphics[width=0.45\textwidth]{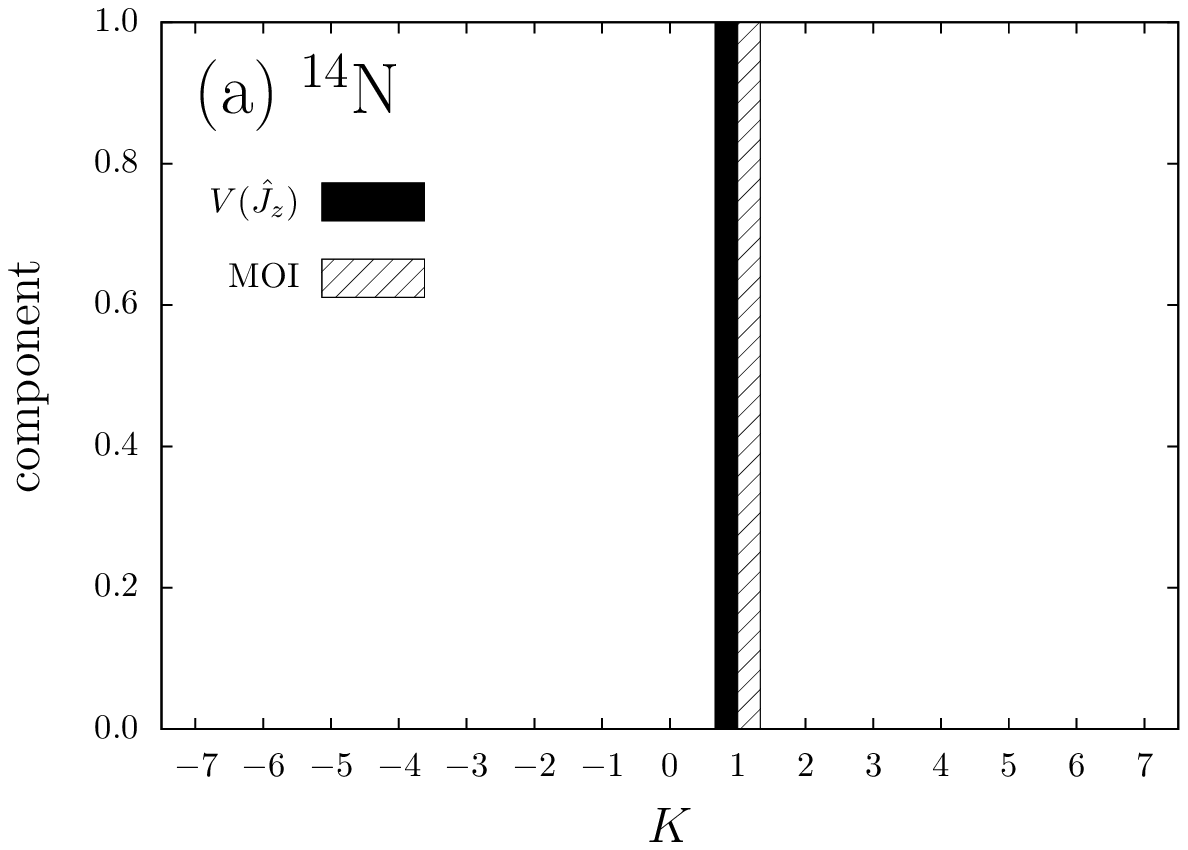}&
   \includegraphics[width=0.45\textwidth]{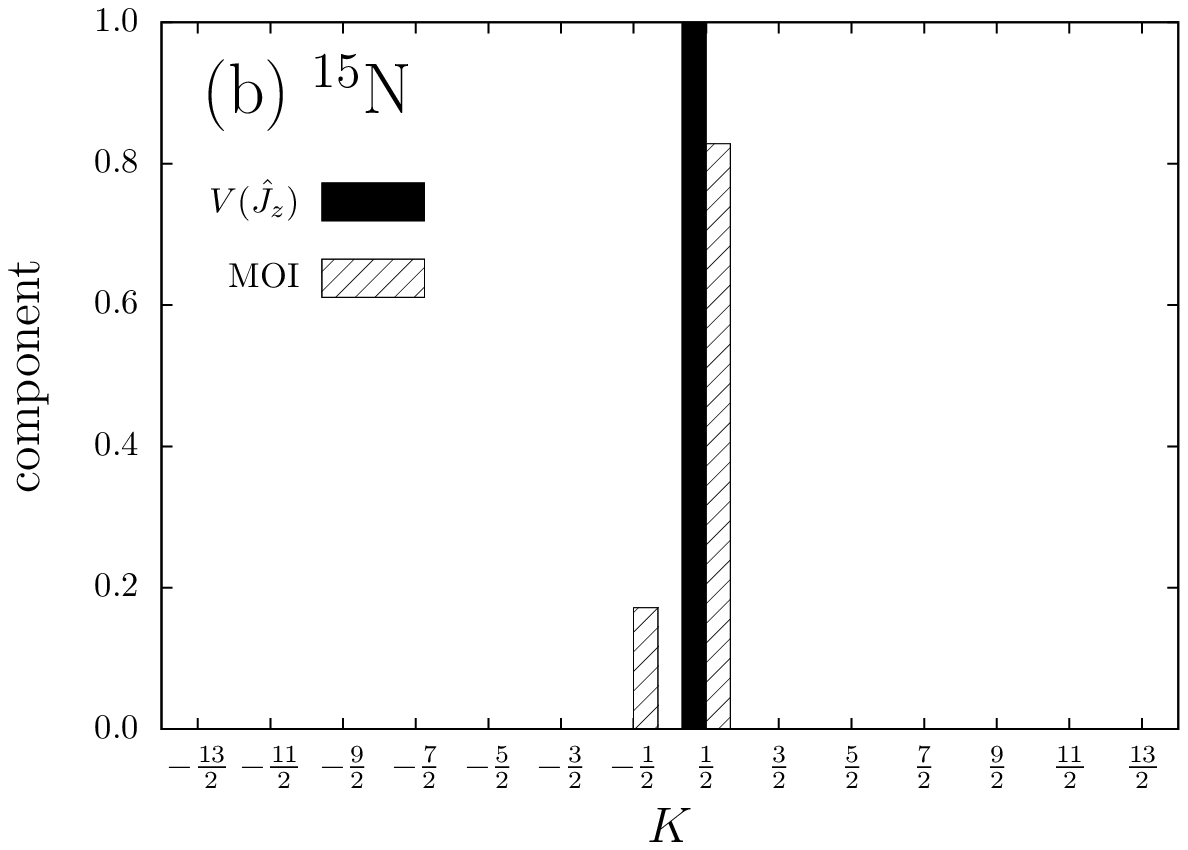}\\
   \includegraphics[width=0.45\textwidth]{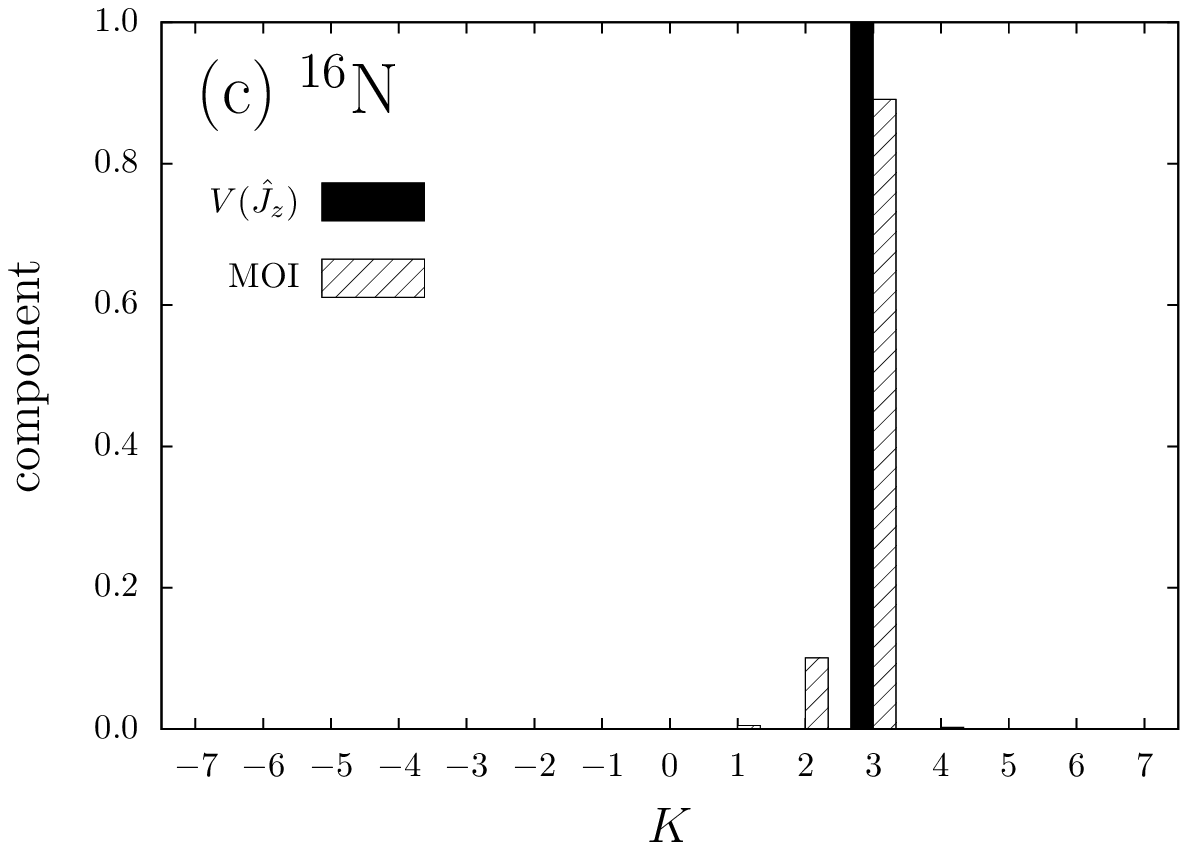}&
  \end{tabular}
  \caption{
  Same as Fig.~\ref{C_K} but for HF states of (a)~$^{14}$N, (b)~$^{15}$N , and (c)~$^{16}$N.
  }
  \label{N_K}
 \end{center}
\end{figure}

Figure~\ref{N_K} shows $K$-distribution for HF states of $^{14,15,16}$N aligned by minimization of variance of $\hat{J}_z$ and diagonalization of moment-of-inertia matrix.
In $^{14}$N, both alignment methods give similar results, and the wave function contain only $K = 1$ components.
In $^{15,16}$N, both methods give different $K$-distribution.
The HF states of $^{15}$N and $^{16}$N aligned by minimization of variance of $\hat{J}_z$ contain only $K = \frac{1}{2}$ and 3 components, respectively.
It shows that those wave functions have axially symmetric form and the the symmetric axes are on each $z$-axis.
Wave functions aligned by diagonalization of moment-of-inertia matrix distributes more than one $K$-components.
$^{15}$N contains $K = \frac{1}{2}$ and $-\frac{1}{2}$ components.
Dominant component of $^{16}$N is $K = 3$ component, and it also contains $K = 2$ components.
They show that symmetric axes of wave functions of $^{15,16}$N aligned by diagonalization of moment-of-inertia matrix deviate from $z$-axes.
Dominant $K$-components of $^{14}$N ($K = 1$) and $^{15}$N ($|K| = \frac{1}{2}$) reflect spin of their ground states that are $J^\pi = 1^+$ and $\frac{1}{2}^-$, respectively.
In $^{16}$N, $K = 3$ components are dominated in the HF state although spin of the ground state in $^{16}$N is $J^\pi = 2^-$.


\begin{figure}[tbp]
 \begin{center}
  \begin{tabular}{cc}
   \includegraphics[width=0.45\textwidth]{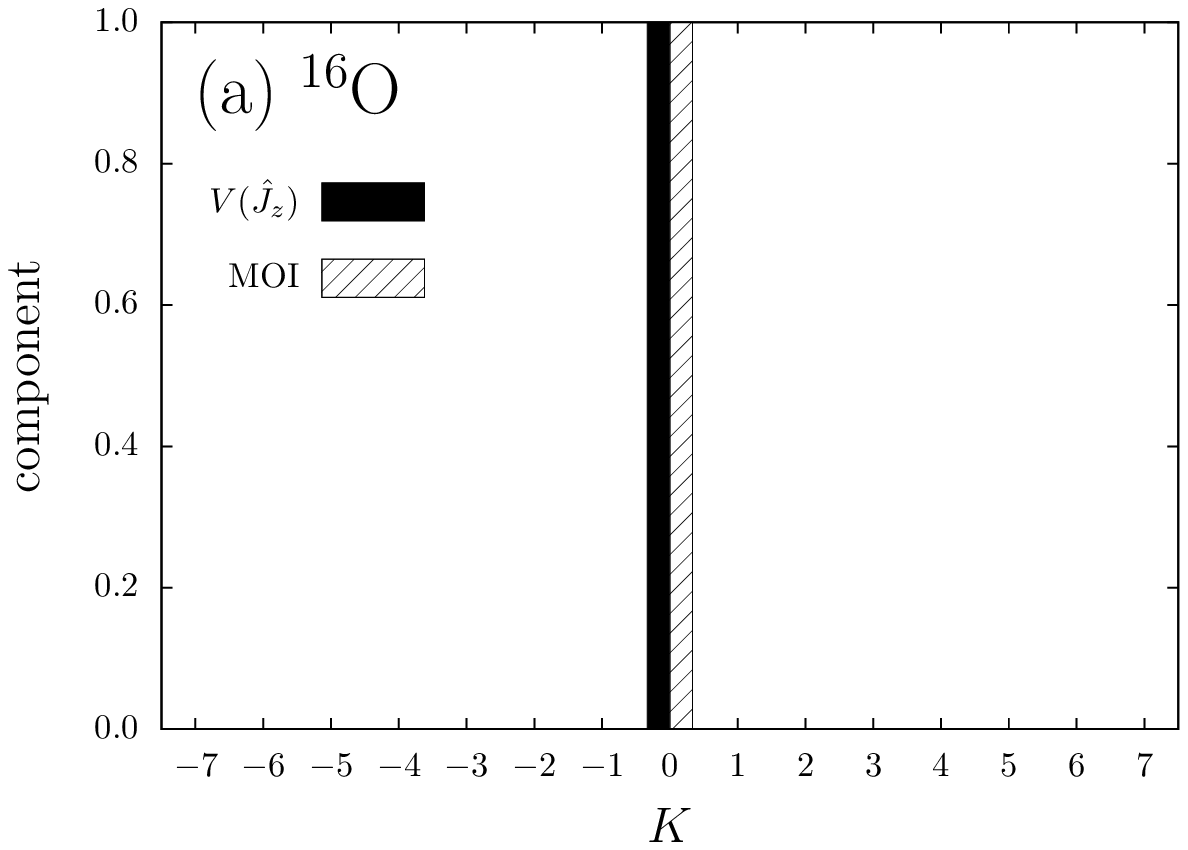}&
   \includegraphics[width=0.45\textwidth]{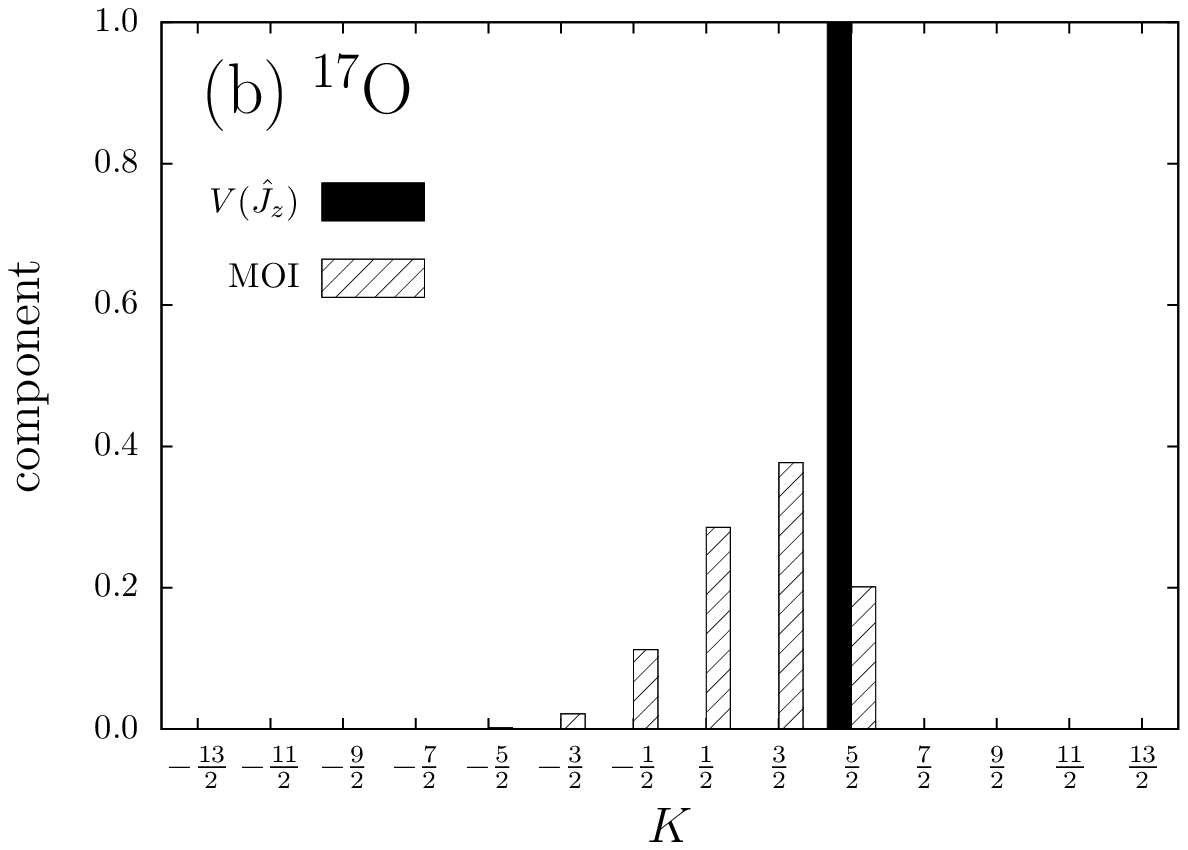}\\
  \end{tabular}
  \caption{
  Same as Fig.~\ref{C_K} but for HF states of (a)~$^{16}$O and (b)~$^{17}$O.
  }
  \label{O_K}
 \end{center}
\end{figure}

Figure~\ref{O_K} shows $K$-distribution for HF states of $^{16,17}$O.
The $K$-distribution of wave functions aligned by minimization of variance of $\hat{J}_z$ and diagonalization of moment-of-inertia matrix are similar in $^{16}$O, which contains only $K = 0$ components.
However, those in $^{17}$O are much different.
The wave function aligned by minimization of variance of $\hat{J}_z$ contains only $K = \frac{5}{2}$ components, which shows that it has axially symmetric form and the symmetric axis is on the $z$-axis.
On the other hand, $K$-components of the wave function aligned by diagonalization of moment-of-inertia matrix are widely distributed from $K = -\frac{3}{2}$ to $K = \frac{5}{2}$.
It shows symmetric axis and $z$-axis are much deviated.
The spin of the ground state in $^{17}$O is $J^\pi = \frac{5}{2}^+$, which reflects dominance of $K = \frac{5}2$ components of the wave function aligned by minimization of variance of $\hat{J_z}$.

\begin{figure}[tbp]
 \begin{center}
  \begin{tabular}{cc}
   \includegraphics[width=0.45\textwidth]{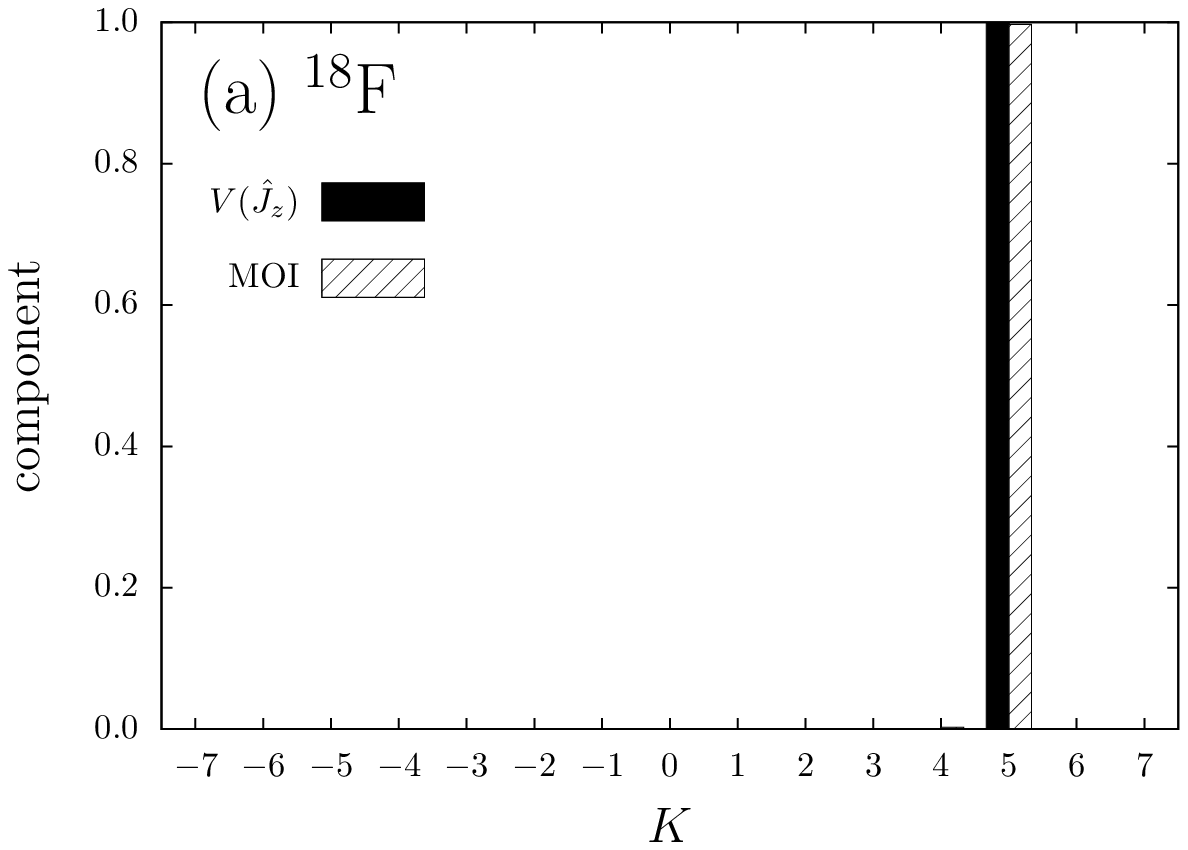}&
   \includegraphics[width=0.45\textwidth]{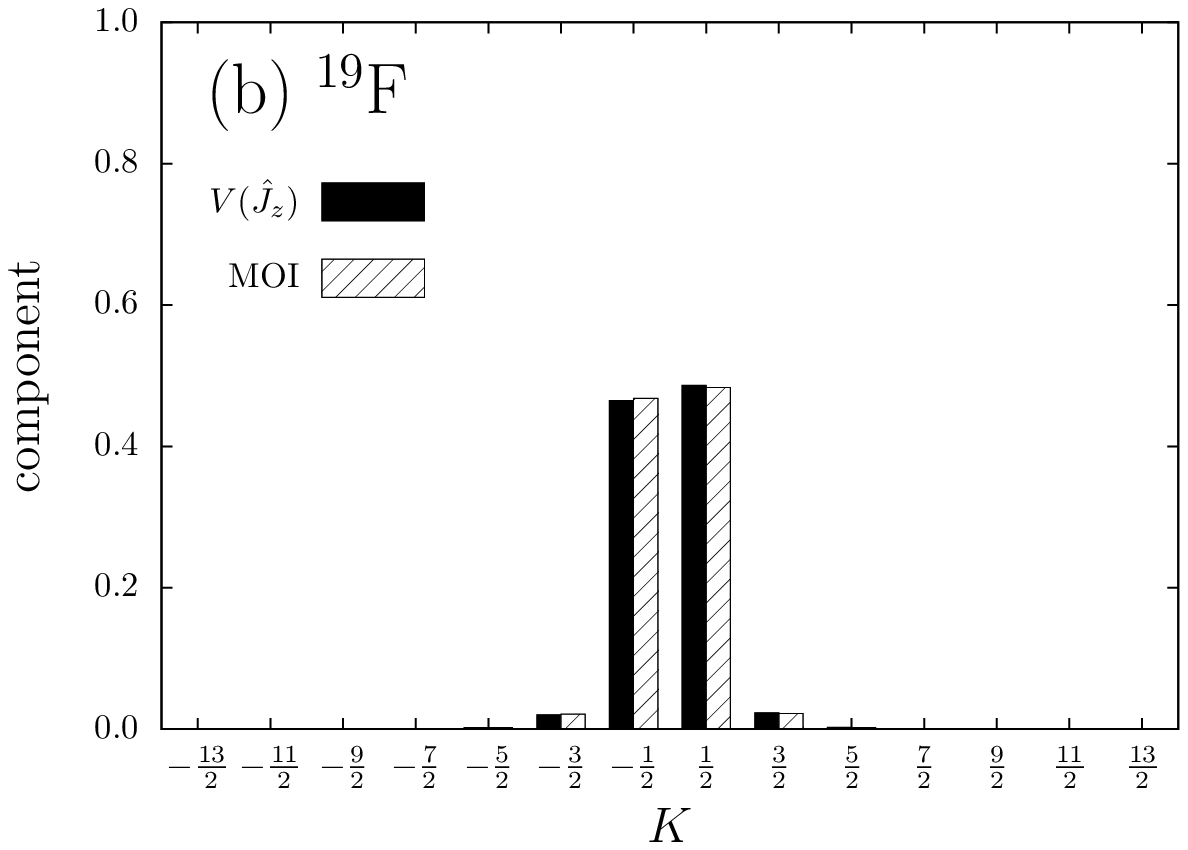}\\
   \includegraphics[width=0.45\textwidth]{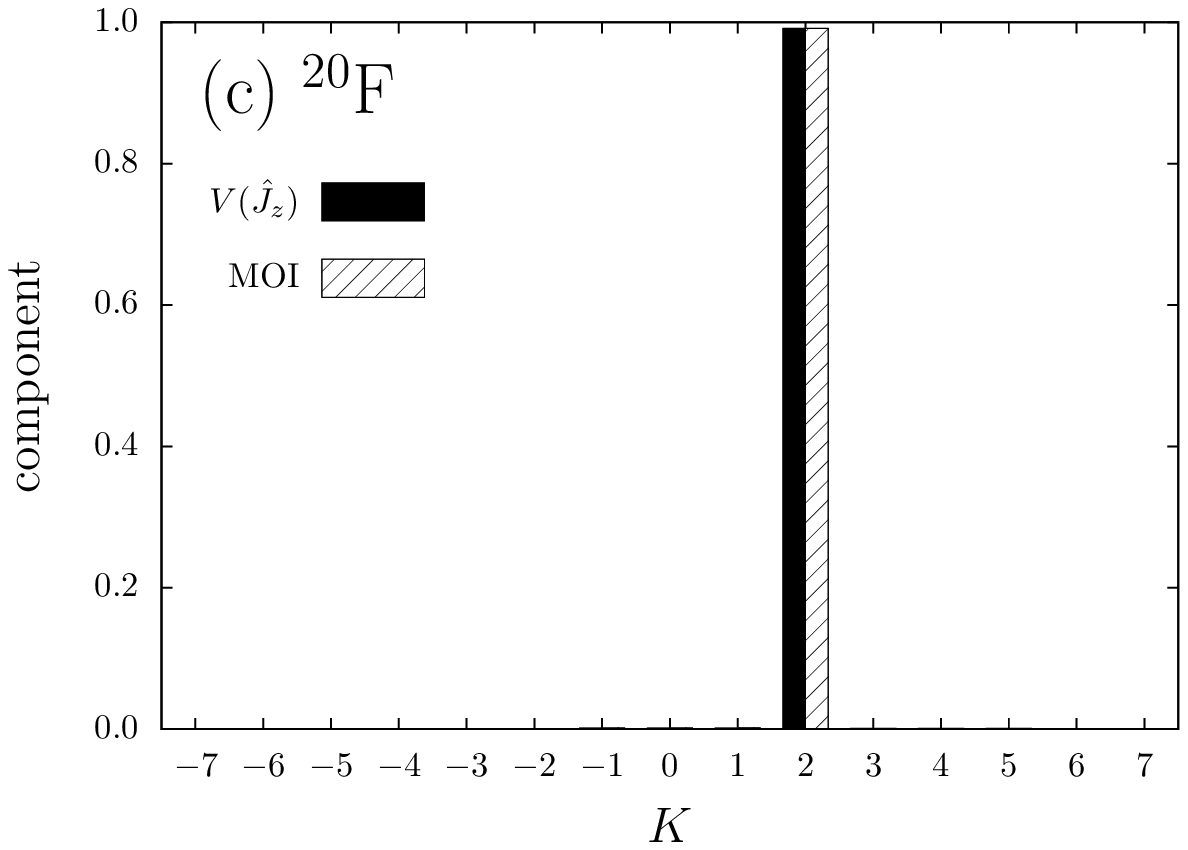}&
  \end{tabular}
  \caption{
  Same as Fig.~\ref{C_K} but for HF states of (a)~$^{18}$F, (b)~$^{19}$F , and (c)~$^{20}$F.
  }
  \label{F_K}
 \end{center}
\end{figure}

\begin{figure}[tbp]
 \begin{center}
  \begin{tabular}{cc}
   \includegraphics[width=0.45\textwidth]{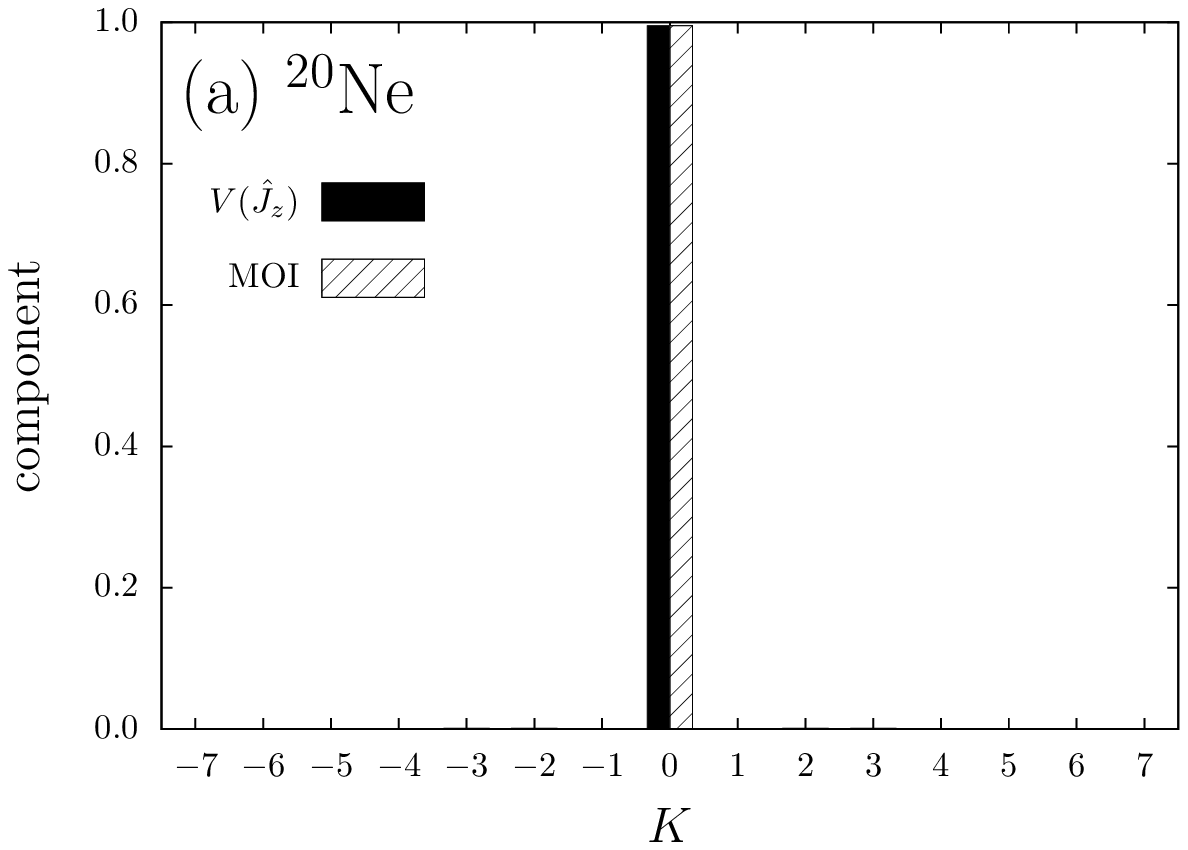}&
   \includegraphics[width=0.45\textwidth]{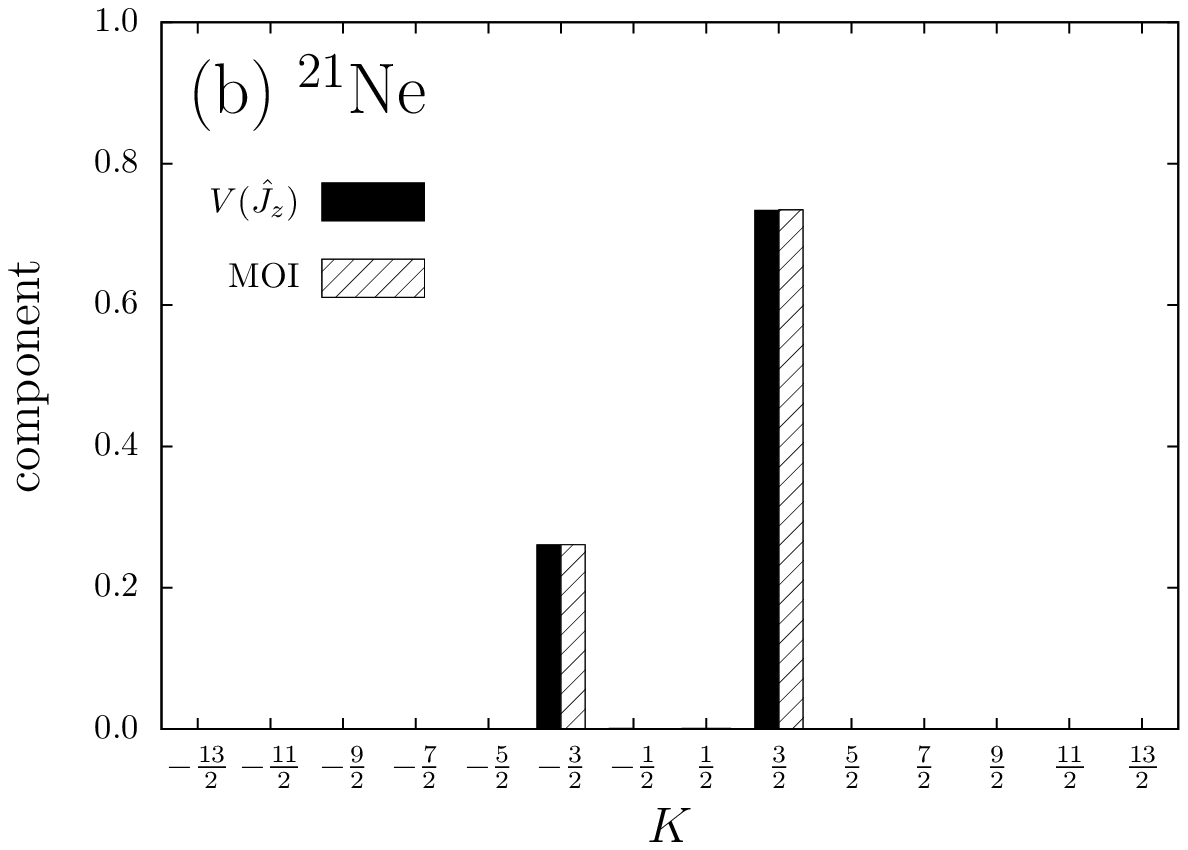}\\
  \end{tabular}
  \caption{
  Same as Fig.~\ref{C_K} but for HF states of (a)~$^{20}$Ne and (b)~$^{21}$Ne.
  }
  \label{Ne_K}
 \end{center}
\end{figure}

Figures~\ref{F_K} and \ref{Ne_K} show $K$-distribution for HF states of $^{18,19,20}$F and $^{20,21}$Ne, respectively, aligned by minimization of variance of $\hat{J}_z$ and diagonalization of moment-of-inertia matrix.
$K$-distribution of wave functions aligned by minimization of variance of $\hat{J}_z$ and diagonalization of moment-of-inertia matrix are similar.
Wave functions of $^{18}$F, $^{20}$F, and $^{20}$Ne contain only $K = 5$, 2, and 0 components, respectively.
Dominant components of $^{19}$F are $K = \pm \frac{1}{2}$ components, and it also contains $K = \pm \frac{3}{2}$ components slightly.
$^{21}$Ne contains $K = \pm \frac{3}{2}$ components.
Dominance of $K = 5$ components in $^{18}$F shows that the wave function does not contain $J = 1$ components although the spin of the ground state of $^{18}$F is $J^\pi = 1^+$.
For $^{19,20}$F and $^{20,21}$Ne, absolute values of dominant $K$-components reflect spin of their ground states, which are $J^\pi = \frac{1}{2}^+$, $2^+$, $0^+$, and $\frac{3}{2}^+$ for $^{19}$F, $^{20}$F, $^{20}$Ne, and $^{21}$Ne, respectively.



\begin{figure}[tbp]
 \begin{center}
  \includegraphics[width=0.45\textwidth]{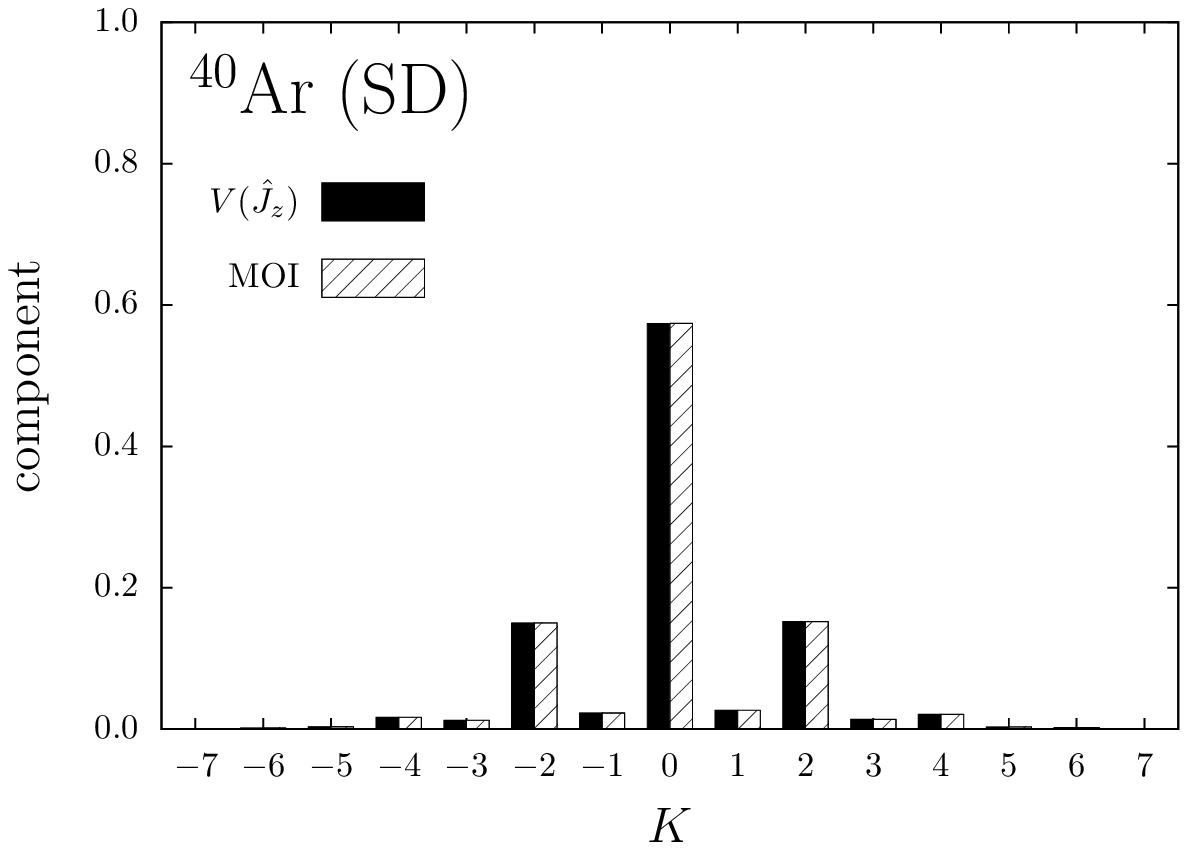}
 \end{center}
 \caption{
 $K$-distribution for triaxially superdeformed states of $^{40}$Ar.
 Filled and slash-pattern bars show expectation values of $K$-projection operator for each $K$ number of wave functions alignment by minimization of variance of $\hat{J}_z$ and diagonalization of moment-of-inertia matrix, respectively.
 }
 \label{Ar_K}
\end{figure}

Figure~\ref{Ar_K} shows $K$-distribution of a triaxially superdeformed state of $^{40}$Ar.
$K$-components of wave functions aligned by minimization of variance of $\hat{J}_z$ and diagonalization of moment-of-inertia matrix are similar.
The wave function contains $K = \pm 2$ components as well as $K = 0$ components, which reflects triaxial deformation of the superdeformed state that has a $K = 2$ side band.
Amounts of $K = +2$ and $-2$ components are similar, which reflects time-reversal symmetry of $^{40}$Ar that is a even-even nuclei.

\section{Discussions}

By comparison of $K$-distribution of wave functions aligned by minimization of variance of $\hat{J}_z$ and diagonalization of moment-of-inertia matrix, it is found that former one has advantages to concentrate $K$-distribution to fewer components, which can improve accuracy of multi-configuration mixing calculation after angular momentum projection.
Typical examples are $\alpha$-$^{24}$Mg cluster structure (Fig.~\ref{Si_K}) and the HF state of $^{17}$O [Fig.~\ref{O_K}(b)].
$K$-distribution of wave functions aligned by minimization of variance of $\hat{J}_z$ and diagonalization of moment-of-inertia matrix are quite different.
Wave functions of $\alpha$-$^{24}$Mg and $^{17}$O aligned by minimization of variance of $\hat{J}_z$ contain only even-$K$ and $K = \frac{5}{2}$ components, respectively.
On the other hand, $K$-components of the wave functions aligned by diagonalization of moment-of-inertia matrix are widely distributed.
The number of contained $K$-components affects accuracy of multi-configuration mixing calculation after angular momentum projection.
When more than one $K$-components are contained in a basis wave function, $K$-mixing is required in a multi-configuration mixing calculation.
The $K$-mixing can make final results worse because different $K$-components are nonorthogonal, which can cause cancellation of digits in diagonalization.
In fact, the wave function of the $^{17}$O aligned by diagonalization of moment-of-inertia matrix gives unphysical energy after angular momentum projection to a $J = \frac{5}{2}$ state with $K$-mixing even when only $K = \frac{1}{2}$ and $\frac{3}{2}$ components, which contain large amounts of $J = \frac{5}{2}$ components, are mixed.
It is because that the $K = \frac{1}{2}$ and $\frac{3}{2}$ components are almost same states after angular momentum projection.
The overlap of the $K = \frac{1}{2}$ and $\frac{3}{2}$ components is 0.9999 after angular momentum projection to $J = \frac{5}{2}$ states.
On the other hand, the wave function aligned by minimization of variance of $\hat{J}_z$ contains only $K = \frac{5}{2}$ components, and $K$-mixing is unnecessary.
Energy after angular momentum projection is safely obtained.

As for $\alpha$-$^{24}$Ar, two alignment methods give different $K$-distribution, and the alignment by minimization of variance of $\hat{J}_z$ works better to concentrate $K$-distribution to fewer components.
$\alpha$-$^{24}$Mg cluster structures have multipole deformed structure, which form $Y_\lambda$ deformation with $\lambda \geq 3$.
Moment-of-inertia is quantity for quadrupole deformation, and it does not reflect multipole deformation although it affects total angular momentum.
Alignment by diagonalization of moment-of-inertia matrix works poorer to concentrate $K$-distribution to fewer components for multipole deformed structures.
In fact, $\alpha$-$^{24}$Mg ($d = 4~\mathrm{fm}$) structure is in oblate side ($\gamma_2 = 42.8^\circ > 30^\circ$), and short axis is set to $z$-axis by diagonalization of moment-of-inertia matrix.
Then density distribution integrated over $z$-axis has neck structure and no symmetry for rotation around the $z$-axis as shown in Fig.~\ref{28Si_density}(b) because the line that connects centers of mass of $\alpha$ and $^{24}$Mg clusters are on the long axis of the cluster structure.
The alignment by minimization of variance of $\hat{J}_z$ selects the long axis as $z$-axis, and the aligned wave function is symmetric for $\pi$ rotation around the $z$-axis [Fig.~\ref{28Si_density}(a)].

As for HF states of $^{15}$N, $^{16}$N, and $^{17}$O, two methods of alignment give different $K$-distribution.
They are  odd or odd-odd nuclei and spherical as quadrupole deformation parameters $\beta_2$ are less than 0.1.
For spherical odd and odd-odd nuclei, $K$-distribution is determined by direction of angular momentum of a proton and/or a neutron in the each highest orbit.
The direction of angular momentum of the highest proton and/or neutron directly reflects variance of $\hat{J}_z$, and by minimization of the variance of $\hat{J}_z$, wave functions are aligned to concentrate  $K$-distribution to fewer components.
On the other hand, diagonalization of moment-of-inertia matrix has no meaning for spherical nuclei, and the method works poorly to concentrate $K$-distribution to fewer components.
Odd and odd-odd nuclei has finite angular momentum even if it is spherical, and by the diagonalization of moment-of-inertia matrix the orientation of angular momentum are set randomly and $K$-distribution can be distributed widely.
In the case of spherical even-even nuclei such as a HF state of $^{16}$O, two methods give similar results because they have only $K = 0$ components.

When wave functions deform largely, two methods give similar results except for multipole deformed structures such as cluster structures.
It is because elements of moment-of-inertia matrix denote orientation of quadrupole deformation, and by diagonalization of moment-of-inertia matrix, wave functions are aligned to concentrate $K$-distribution to fewer components.

The HF states of $^{16}$N and $^{18}$F contain only $K = 3$ and 5 components, respectively.
The $K$ numbers are larger than spin of their ground states, which are $J^\pi = 2^-$ and $1^+$ for $^{16}$N and $^{18}$F, respectively.
It shows that components of their ground states are not contained in their HF states.
In low-lying states of $^{16}$N and $^{18}$F, $J^\pi = 3^-$ (0.298~MeV) and $5^+$ (1.121~MeV) states exist, respectively, and dominant components of those low-lying excited states are obtained as the HF state in simple mean-field calculation.
In order to obtain dominant components of $J^\pi = 2^-$ and $1^+$ ground states in $^{16}$N and $^{18}$F, respectively, more complicated calculation are necessary such as energy variational calculation with a constraint.
The HF states of $^{16}$N and $^{18}$F deform oblately in the current framework, and their prolate states can be dominant components of their ground states.

\section{Conclusions}

Minimization of variance of $\hat{J}_z$ is proposed as a useful method to align wave functions for angular momentum projection.
Advantages of the proposed method are shown by benchmark calculations using the AMD wave functions of $\alpha$-$^{24}$Mg structure of $^{28}$Si, triaxially SD states of $^{40}$Ar, and HF states of $^{12,13}$C, $^{14,15,16}$N, $^{16,17}$O, $^{18,19,20}$F, and $^{20,21}$Ne.
Comparing with wave functions aligned by diagonalization of moment-of-inertia matrix, $K$-distribution of wave functions aligned by the proposed method is more concentrated, which has advantages for accurate multi-configuration mixing calculation with angular momentum projection.
Especially, the proposed method works well for multipole deformed structures such as cluster structures, and spherical structures of odd and odd-odd nuclei.
For largely quadrupole deformed structure of even-even nuclei, two methods give similar results.

\section*{Acknowledgment}

The author thanks to Dr.~Kimura and Mr.~Chiba in Hokkaido University for fruitful discussions.
Numerical calculations were conducted on COMA in Center for Computational Sciences, University of Tsukuba.
This work was supported by JSPS KAKENHI No.~25800124.

\bibliographystyle{ptephy}
\bibliography{alignment_v2}
%

\end{document}